\documentclass[aps,twocolumn,showpacs,superscriptaddress,
amsmath,amssymb,tightenlines]{revtex4}

\usepackage{graphicx} 
 
\begin{document} 
 
\title{Optomechanical cooling  of levitated spheres with doubly-resonant fields}

\author{G A T Pender}
\affiliation{Department of Physics and Astronomy,
University College London, Gower Street, London WC1E 6BT, United Kingdom}

\author{P F Barker}
\affiliation{Department of Physics and Astronomy,
University College London, Gower Street, London WC1E 6BT, United Kingdom}

\author{Florian Marquardt}
\affiliation{Institut for Theoretical Physics, Universit\"at Erlangen-N\"urnberg, Staudtstra\ss e 7, 91058 Erlangen Germany}

\author{James Millen}
\affiliation{Department of Physics and Astronomy,
University College London, Gower Street, London WC1E 6BT, United Kingdom}

\author{T S Monteiro}
\affiliation{Department of Physics and Astronomy,
University College London, Gower Street, London WC1E 6BT, United Kingdom}

\begin{abstract}
Optomechanical cooling of levitated dielectric particles represents a promising
new approach in the quest to cool small mechanical resonators towards their
quantum ground state.  We investigate two-mode cooling of levitated nanospheres
in a self-trapping regime. We identify a rich structure of split sidebands
(by a mechanism unrelated to usual strong-coupling effects) and strong 
cooling even when one mode is blue detuned.
We show the best regimes occur when both optical fields cooperatively cool
 and trap the nanosphere, where cooling rates are over an order of magnitude faster
compared to corresponding single-sideband cooling rates. 
\end{abstract} 

\maketitle

Extraordinary progress has been made in the last half-dozen years \cite{Review,Kipp}
towards the final goal of cooling a small mechanical resonator down to its
 quantum ground state and hence to realise quantum behavior
in a macroscopic system. Implementations include cavity cooling of micromirrors on
 cantilevers \cite{Metzger,Arcizet,Gigan,Regal},
dielectric membranes in Fabry Perot cavities \cite{membrane};
 radial and whispering gallery modes of optical microcavities \cite{Schliess}
 and nano-electromechanical systems \cite{NEMS}. Indeed the realizations
span 12 orders of magnitude \cite{Kipp},  up to and including the LIGO gravity wave experiments.
Corresponding advances in the theory of optomechanical cooling have also been
made \cite{Brag,Paternostro,Marquardt,Wilson}.

Over the last year or so, a promising new paradigm has been attracting much
interest: several groups \cite{Isart,Zoller,Barker,Ritsch} have now proposed schemes 
for optomechanical cooling of
levitated dielectric particles, including nanospheres and even viruses \cite{Isart}. 
The important advantage is the elimination of the
mechanical support, a dominant source of heating noise.
In general, these proposals involve two fields, one for trapping and one for cooling.
This may involve an optical cavity mode plus a separate trap; or two optical cavity modes,
the so-called ``self-trapping'' scenario.

 Mechanical oscillators in the
self-trapping regime differ from other optomechanically-cooled devices in a second
 fundamental respect (in addition to the absence of mechanical support): the mechanical frequency, 
$\omega_M$, associated
 with  centre of mass oscillations is not an intrinsic feature of the
resonator but is determined by the optical field. In particular, it is a function of one or
both of the detuning frequencies, $\delta_1$ and $\delta_2$, of the optical modes. 
Cooling, in general, occurs when $\omega_M$ is resonantly red detuned with either of the detuning frequencies (i.e. negative $\delta_{1,2}$ is associated with cooling). 
For self-trapping systems, this means
 $\omega_M(\delta_1,\delta_2) \sim -\delta_{1,2}$ so the relevant frequencies are not
independent. 

 The full implications of
this nonlinear interdependence of the resonant frequencies have not yet been fully elucidated.
We show here for the first time that it leads to a rich landscape of split sidebands.
The  mechanism here is unrelated to splittings
seen in experiments in the strong-coupling regime \cite{strong}. However, it
results in extremely favourable cooling regimes, where two (or more) cooling
sidebands approach each other. We term this the ``double-resonance'' regime.
 We find it can produce cooling rates nearly two orders of magnitude
stronger than the corresponding ``single-resonance'' case.
\begin{figure}[h] 
\begin{center}
 \includegraphics[width=3in]{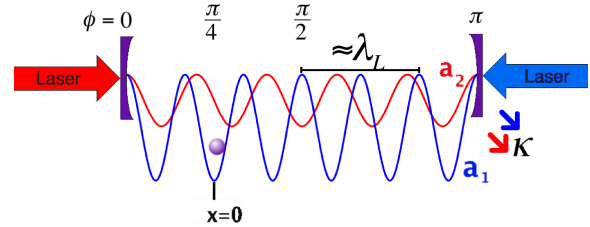}
\end{center}
\caption{Schematic set-up: a levitated nanosphere is trapped and cooled cooperatively
by two optical modes. The optical potential for each mode is shown.} 
\label{Fig0} 
\end{figure} 
A self-trapping Hamiltonian was investigated in \cite{Zoller} and corresponds to
the set-up illustrated in Fig.\ref{Fig0}: 
\begin{eqnarray} 
\hat H&=& -\delta_1{\hat a}_1^\dagger {\hat a}_1  -\delta_2 {\hat a_2}^\dagger {\hat a}_2 
      +  \frac{{\hat P}^2}{2m}   -A  {\hat a}_2^\dagger  {\hat a}_2 \cos^2 (k_2 x-\phi) 
         \nonumber \\
 &- & A{\hat a}_1^\dagger {\hat a}_1\cos^2 k_1 x
  + E_1({\hat a_1}^\dagger+ {\hat a}_1) + RE_1({\hat a}_2^\dagger + {\hat a}_2)
\label{Hamiltonian}
\end{eqnarray}
 Two optical field modes ${\hat a}_{1,2}$ are coupled to a nanosphere with centre of mass position $x$.
$\hat H$ is given in the rotating frame of the laser which drives the modes with amplitudes
$E_1$ and $R E_1$ respectively. In \cite{Zoller}, the phase between the optical potentials
was chosen to be $\phi=\pi/4$; the study focussed primarily on the $\delta_1\simeq 0$ regime, where 
the ${\hat a}_{1}$ mode is responsible exclusively for trapping
 while the ${\hat a}_{2}$ mode alone provides cooling.
Previous studies \cite{Isart,Zoller,Ritsch} all analysed mechanical oscillations about an
equilibrium position $x_0 \simeq 0$, corresponding to the antinode of the trapping mode
(field 1).
Below, this scenario is referred to as the ``single-resonance'' regime.

Here we investigate the effects of relaxing all these restrictions and
find interesting and unexpected implications.
We take $\phi=\pi/4$; the cooling field is driven more weakly than the 
trapping field, but with a ratio $R \simeq 0.1-1$. Below,
our analytical expressions cover arbitrary $\kappa,R,E_1,A$, but we compare with
 an illustrative set of { \em experimentally plausible parameters}: we take a 
 cavity damping ${\kappa} =6\times 10^{5}$Hz. We considered driving powers in the
range $P \simeq 1-10$ mW, where $P=\frac{2kc E_1^2 \hbar}{\kappa}$.
For a laser of wavelength $\lambda=1064$ nm and a cavity of length $L \sim 1$ cm,
waist $25\mu m$
we consider a silica nanosphere of $100$nm radius and hence a coupling strength 
$A\simeq 3\times 10^{5}$ Hz. To obtain a dephasing of $\pi/4$ between the two modes near the centre of the cavity, the frequency
difference between the modes is $|\omega_1-\omega_2| \sim 2\pi \times 10$ GHz.
This far exceeds the detunings $\delta_{1,2} \sim 1$ MHz
and  also the mechanical frequencies $\omega_M$. Thus the photons are 
completely distinguishable and can be read out and driven separately.
Nevertheless, since $\omega_{1,2} \sim 10^{14}$ Hz, we
approximate $k_1 \simeq k_2 \equiv k$. However, this situation is distinct from
the ring-cavity proposal of 
 \cite{Ritsch} where $R=0$ and mode 2 is undriven but is populated exclusively
by scattering from mode 1; there, the photons are of precisely the same frequency 
and thus there is  
 a single detuning parameter involved.

\begin{figure}[ht]
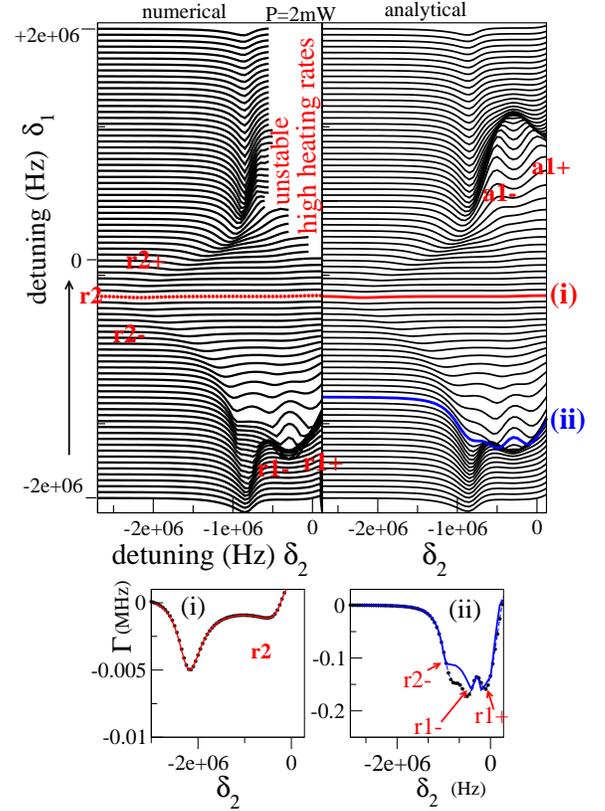
 
\begin{center}

 \includegraphics[width=3.in]{Fig1a.eps}
 \includegraphics[height=1.2in]{Fig1b.eps} 
\end{center}
\caption{(Colour online){\bf Upper Panels:} comparison between numerical optical cooling rates
(without linearisation) and an analytical 
expression (Eq.\ref{gamma}) from linearised dynamics, showing excellent agreement. $R=0.5$.
The curves corresponding to different values of $\delta_1\pm 2$MHz
are shifted relative to each other. 
At single-resonance $r2$, field 2 is resonant with the oscillator and is exclusively responsible for cooling;
field 1 is resonant with the cavity ($\Delta_1^x=0$) and traps the sphere.
Subsequently $r2\pm$ appear: they are cooling resonances of field 2, split by field 1;
  conversely  $r1\pm$ are cooling resonances of field 1, split by field 2. 
  $r2-,r1-,r1+$ overlap giving a broad region
 of very strong cooling. The
 $a1\pm$ are heating (Stokes) resonances of field 1. The $a1-$ can coincide with
the cooling resonance ($r2+$). Here field 2 absorbs phonons as fast as field 1 emits
them. 
{\bf Lower Panels:} Show the unshifted cooling curves at the single-field cooling resonance $r2$
and double-resonant cooling, showing that the latter gives over an
order of magnitude stronger cooling. Asterisks are numerical results, blue and
red curves correspond to curves in upper panels.}
\label{Fig1} 
\end{figure}

Fig.\ref{Fig1} illustrates the behavior for $R=0.5$. It
shows that allowing both fields to cooperatively
trap and cool yields more than an additive improvement.
 We denote by $r2$ and $r1$ the set of detunings 
corresponding to cooling resonances of 
modes 1 and 2 respectively: Fig.\ref{Fig1} shows that these resonances 
unexpectedly split and separate into new cooling resonances $r1\pm$
and $r2\pm$. These can overlap, to give very strong cooling associated with
multiple resonances.  Below, we also show that the usually studied
single-field resonance regime $r2$ can attain only a maximal cooling
 rate $\Gamma \simeq \frac{R^2 A}{4}$;
and that for strong driving, $ \Gamma \propto 1/E_1$: thus, the cooling falls with 
increasing driving and it is hard to achieve optimal cooling regimes where
$\Gamma \simeq A \sim \kappa$.
For the double-field resonances, in contrast, $\Gamma \propto E_1^{1/2}$,
the cooling increases with $E_1$ and can more easily reach optimal cooling.
 Very strong cooling is apparent even for regimes where one mode
is blue detuned.
In addition, although there is no direct coupling
between optical modes, double-resonances offer the prospect of strong (albeit second-order)
 coupling and entangling of the two modes via
the nanosphere, within a single cavity. This includes simultaneous resonant/antiresonant
regimes (indicated by the crossing of $r2+$ and $a1-$ in Fig.\ref{Fig1})
where one mode resonantly heats, while the other resonantly cools the mechanical
mode.

The dynamics depends on $k,m,A,\kappa,\delta_1,\delta_2,E_1,R$.
 However transforming to scaled variables reduces this complexity. We rescale
 position, time and field variables as follows:
$kx \to {\tilde x}$, $At \to {\tilde t}$, then  $a_{1,2} \to \frac{E_1}{iA} {\tilde a_{1,2}}$ 
 Note that below we drop all the tildes but it is implicit that all variables are scaled in the resulting Heisenberg equations:
\begin{eqnarray}
{\ddot {\hat x}} & = & -\epsilon^2 \left[|{\hat a}_1|^2 \sin 2x + |{\hat a}_2|^2 \sin (2x-\pi/2)\right] \nonumber\\ 
{\dot {\hat a}_1}& = & i \Delta_1 {\hat a}_1 +1 +i {\hat a}_1\cos^2 x -\kappa_A {\hat a}_1 \nonumber \\
{\dot {\hat a}_2} & = & i \Delta_2 {\hat a}_2 +R +i {\hat a}_2\cos^2 (x-\pi/4) -\kappa_A {\hat a}_2 .
\label{a2} 
\end{eqnarray}
The dynamics for a given $R<1$ depends only on the
the scaled driving $\epsilon^2= \zeta E_1^2$ where $\zeta=\frac{\hbar k^2}{m A^3}$,
 two scaled detunings $\Delta_{1,2}= \delta_{1,2}/A$ and
a scaled damping $\kappa_A=\frac{(\kappa/2)}{A}$; all scaled frequencies (including
cooling rates) are given below as a fraction of $A$.

The experimentally adjustable parameters are $\epsilon$, both the detunings $ \Delta_{1,2}$ and $R$.
We assume $\kappa_A \simeq 1$, though the analytical expressions are for arbitrary
$\kappa_A$. Varying driving power $\mathcal{P} \sim 1-10$ mW , but leaving the cavity/nanosphere properties
unchanged means $A$ remains constant, but $\epsilon^2$ varies from $\sim 1-100$. 

Following the usual procedure, we replace operators by their expectation values and
 linearise about equilibrium fields by performing the shifts $a_1 \to \alpha_1 + a_1$,
$a_2 \to \alpha_2 + a_2$ and $x \to x_0 + x$.
Hence we find equilibrium photon fields, 
$\alpha_{1} =  \left[\kappa_A - i\Delta_{1}^x \right]^{-1} \ $ and
$\alpha_{2}  =  R\left[\kappa_A - i\Delta_{2}^x \right]^{-1}$ as well as
position $\tan 2x_0 = {|\alpha_2|^2}/{|\alpha_1|^2}$.

Here, $\Delta_1^x=\Delta_1+\frac{1}{2}(1 +\cos 2x_0)$ and 
 $\Delta_2^x=\Delta_1+\frac{1}{2}(1 +\sin 2x_0)$.
The dimensionless mechanical frequency is:
\begin{eqnarray}
\omega_M^2(\Delta_1,\Delta_2)= 2\epsilon^2(|\alpha_1|^2 \cos{2x_0}  + |\alpha_2|^2 \sin{2x_0} ).
\label{freqs}
\end{eqnarray}
Closely related forms of this ``self-trapping'' frequency expression
 have been noted previously \cite{Isart,Zoller,Barker,Ritsch} but
the implications, other than for $x_0 \simeq 0$, have not been investigated. 


 To first order, the linearised equations of motion are:
\begin{eqnarray}
{\ddot x}& = & - \omega^2_M x - \epsilon^2(g_1\sin 2x_0-g_2\cos 2x_0 )\nonumber\\
{\dot a_1}& = & i \Delta_1^x a_1 - i  \alpha_1 x \sin 2x_0  -\kappa_A a_1 \nonumber\\
{\dot a_2} & = & i \Delta_2^x a_2 + i \alpha_2 x \cos 2x_0 -\kappa_A a_2
\label{a2s} 
\end{eqnarray}
where $g_i=(\alpha_i^* a_i+ \alpha_i a_i^*)$. 
From the above, we can obtain the contribution from the two photon fields to 
the optomechanical cooling:
\begin{equation}
\frac{\Gamma}{2}=\frac{\epsilon^2\kappa_A}{2\omega_M} 
\left[S_1(\omega_M)+S_2(\omega_M)-S_1(-\omega_M)-S_2(-\omega_M)\right]
\label{gamma}
\end{equation}
where 
\begin{equation}
S_1(\omega)=\frac{|\alpha_1|^2\sin^2 2x_0 }{[\Delta_1^x-\omega]^2 + \kappa_A^2}; \ 
S_2(\omega)=\frac{ |\alpha_2|^2\cos^2 2x_0}{[\Delta_2^x-\omega]^2 + \kappa_A^2}
\label{gamma12}
\end{equation}
(net cooling occurs for $\Gamma <0$).
We also calculate a numerical $\Gamma $ by evolving the
equations of motion in time and looking at the decay in $x(t)$
(its variance in particular).
The analytical cooling rates give excellent agreement with numerics in all but
the strongest cooling regions. 

The single-field cooling resonance $r2$ occurs 
for $x_0 \simeq 0$,  $\Delta_1^x \simeq 0$ and $\Delta_2^x=-\omega_M$,
thus here for $\Delta_1=-1$ (i.e. $\delta_1=-A$).
Conversely, there is also a single-field cooling resonance $r1$ for 
$ x_0 = \pi/4$ (note that since we consider only the case $R<1$, i.e. field 2
is always driven more weakly than field 1, the latter situation does not correspond simply
to an interchange in the role of the fields).
Away from these extreme cases, both cooling resonances
are split by the effect of the other field, wherever $0 < x_0 < \pi/4$.

From Fig.\ref{Fig1} we see
 $r2\pm$ occur for the same $\Delta_2$, thus the same equilibrium
photon field $\alpha_2$; however they correspond to photon fields
$\alpha_1$ and $\alpha_1^*$ respectively and thus to 
different $\Delta_1^\pm= \pm y_1$, where $2 y_1 $ is the 
splitting (about $\Delta_1^x=0$)
between $r2+$ and $r2-$ seen in Fig.\ref{Fig1}.

 The transformation $\alpha_1 \to \alpha_1^*$
leaves both the mechanical frequency $\omega_m(\Delta_1^\pm,\Delta_2)$
 and $x_0(\Delta_1^\pm,\Delta_2)$ unchanged.
Hence the cooling rates are similar for both $r2\pm$.

We can estimate the splitting $\Delta_1^\pm$ by requiring
\begin{equation}
\omega_M(\Delta_1^+,\Delta_2)=\omega_M(\Delta_1^-,\Delta_2)\simeq \Delta_2^x
\end{equation}
since  $\omega_M(\Delta_1^\pm,\Delta_2)\simeq \Delta_2^x$ are the conditions for
the optomechanical resonance $r2\pm$. From Eqs.(\ref{aeq}) and (\ref{freqs})
we see:
 \begin{equation}
 \pm y_1=\pm \sqrt{\frac{2\epsilon^2}{(\Delta^x_2)^2 \cos 2x_0} -\kappa_A^2 }.
\end{equation}
Close to $r2$, we can simplify
$y_1 \simeq \pm \frac{\sqrt{2}\epsilon}{\Delta_2+A/2}$.
Similarly, $y_2$, the splitting between $r1\pm$ is
$\pm  y_2=\pm \sqrt{\frac{2\epsilon^2R^2}{(\Delta_1^x)^2 \sin 2x_0} -\kappa_A^2 }$.

\begin{figure}[htb] 
\begin{center}
 \includegraphics[height=3.2in]{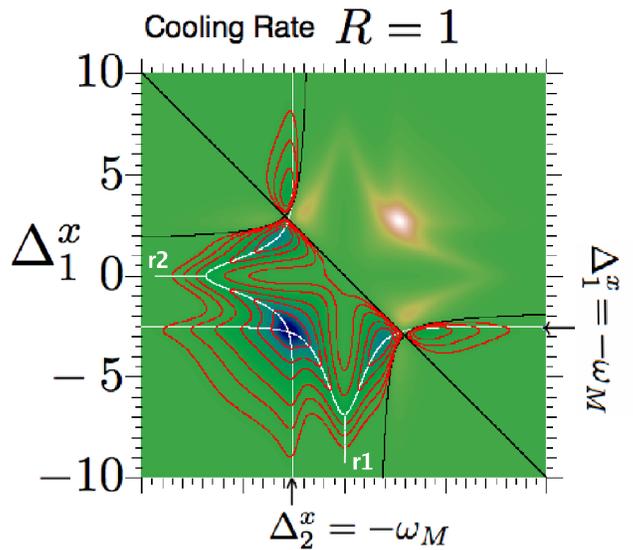} 
\end{center}
\caption{(Colour online) Cooling rates as a function of scaled detunings.
For $R=1$ the behaviour of mode 1 and mode 2 is equivalent, thus a high degree
of symmetry is evident. The splitting of the resonances is much larger, but
very strong cooling maximum (dark blue) at the double-resonance of $r2-,r1-$ is seen, with a matching
heating maximum (white/orange) for $\Delta_{1,2}^x >0$.}
\label{Fig2} 
\end{figure}

Thus the splittings increase with driving power and $R$. While in Fig.\ref{Fig1},
corresponding to $R=0.5$, three resonances ($r2-, r1-$ and $r1+$) overlap, 
 in  Fig.\ref{Fig2}, for $R=1$ 
 $r1\pm$ are well separated and the double resonance involves only $r2-,r1-$.

\begin{figure}[htb] 
\begin{center}
 \includegraphics[height=2.in]{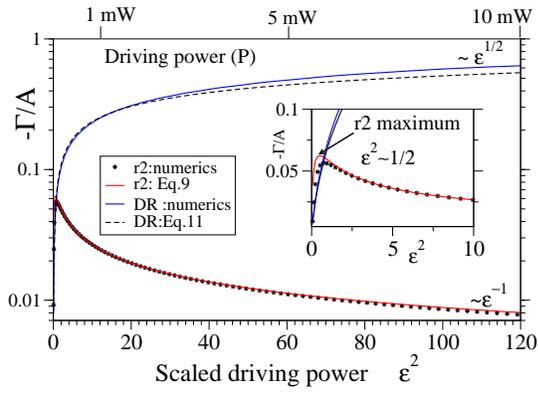} 
\end{center}
\caption{Comparison between single-resonant ($r2$) and double
resonant (DR) i.e. simultaneous $r1-$ and $r2-$ resonant 
cooling rates as a function of laser driving power ($R=0.5$),
showing that the double resonance corresponds to two orders of magnitude greater cooling
for strong driving. Inset shows cooling maximum of $r2$
at weak driving $\epsilon^2 \simeq 1/2$; here the system is on
the edge of the sideband-resolved regime $\omega_M=\kappa_A$. 
In contrast for DR, at $P=10mW$, $\omega_M/\kappa_A \simeq 4$ and the
$r1-$ and $r2-$ sidebands are very well resolved.}
\label{Fig3} 
\end{figure}

We now analyse the relative merits of single-resonance versus
double-resonance cooling. Single-field cooling corresponds to $r2$ in Fig.\ref{Fig1}.
 Cooling rates are obtained from
Eq.\ref{gamma} by taking $x_0 \simeq 0$,  $\Delta_1^x=0$ and $\Delta_2^x=-\omega_M$.
This regime was investigated in \cite{Zoller} and Eq.\ref{gamma} reduces to
expressions therein (in unscaled units). 
However, we can give good approximations to cooling rates purely in terms of
experimental parameters (driving power, $R$ and $\kappa$).
Assuming $S_2(-\omega_M)\gg S_2(+\omega_M)$ and that the field 1 contribution to cooling
is negligible,
near $r2$, the mechanical frequency $\omega_M^2=  \frac{2\epsilon^2}{\kappa_A^2}$.
Hence, as shown in the appendix, the Single Resonance (SR) cooling rate becomes:
\begin{equation}
-\Gamma_{SR}\approx \frac{R^2\epsilon\kappa_A^2}{\sqrt{2}}(2\epsilon^2+\kappa_A^4)^{-1}
\label{SRA}
\end{equation}
(recall this is a scaled cooling rate thus given in units of $A$).

Single-field cooling is a maximum if $\epsilon=\kappa_A^2/\sqrt{2}$ where
$\omega_M=\kappa/2$ (in unscaled units) and is thus at the edge of the resolved sideband regime.
Here, $-\Gamma_{SR}\approx \frac{R^2}{4}$; this gives optimal cooling $\Gamma \sim \kappa$
 only if $R \sim 1$. This cooling maximum is independent of $\kappa_A$: it depends only on $R$.
 As the driving is increased, if $2\epsilon^2 \gg \kappa_A^4$, 
 \begin{equation}
-\Gamma_{SR}\sim \frac{R^2 \kappa_A^2}{2\sqrt{2} \epsilon} \propto 1/\epsilon
\label{SRRE}
\end{equation}
Thus the single resonance cooling rate falls off quite rapidly as the driving
amplitude is increased: the  cooling cannot be improved 
by increasing the driving amplitude.

To obtain the corresponding double-resonant rate (DR) one must first
identify the $\epsilon$-dependent 
pair of detunings for which 
$-\omega_M(\Delta_1,\Delta_2) \approx \Delta^x_1 \approx \Delta^x_2$.
Even if $r1-,r2-$ do not cross in Fig.\ref{Fig1}, the sidebands approach 
within their width $\kappa$ and overlap significantly. One can still obtain a good
approximation to the cooling rate in terms of driving parameters (see Appendix).
In this case, there are  contributions to cooling from both field 1 and field 2. Adding
them both,
\begin{equation}
-\Gamma_{DR}\approx \frac{\epsilon^2 (R^2+R^4)}{ (\kappa_A \omega_M (\omega_M^2+\kappa_A^2)}
\label{DRA}
\end{equation}
where the frequency is given by the expression,
 $\omega_M^2=\frac{-\kappa_A^2}{2} + \frac{1}{2} \sqrt{ \kappa_A^4+ 8\epsilon^2} $.
The contribution from mode 2 $\propto R^2$ while that of mode 1 $\propto R^4$; both contribute
significantly for $R \gtrsim 0.5$.
Assuming $\omega_M \gg \kappa_A$, this reduces to
$-\Gamma_{DR}\approx 2^{-3/4}{(R^2+R^4)\epsilon^{1/2}}/\kappa_A$ , hence $\Gamma_{DR} \propto \epsilon^{1/2}$.

Fig.\ref{Fig3} shows that Eqs.\ref{SRA} and Eqs.\ref{DRA} both give excellent agreement with
exact numerics.  In the double-resonant case, the cooling is stronger and 
increases with increasing $\epsilon$.  In contrast, the single-resonant cooling
 $\Gamma_{SR} \propto 1/\epsilon$
and this cannot be improved by increasing $\epsilon$. Self-trapping cooling
cannot be considered simply in terms of an additive contribution from two intracavity 
intensities in Eq.\ref{gamma12}; the response of $\omega_M$ to the driving is also
important. In the double-resonant case,
 $\omega_M\propto \epsilon^{1/2}$ for strong driving. In contrast, for 
the single-resonant case $\omega_M \propto \epsilon$ and strong driving pushes the
  $r2$ resonance into the far-detuned regime. 
A study of the quantum cooling shows that the minimum phonon numbers attainable
$\bar{n}_{min} \ll 1$ for single and double resonant cooling for strong driving
$\epsilon^2 \gg 1$. For weak driving, strong cooling can be obtained with
single field cooling, but at the edge of the resolved sideband regime $\omega_M=\kappa/2$,
less favourable for ground state cooling. \\
{\em Conclusion} 
Our study shows that the two-mode self-trapping regime has a raft of
of unexpected features, including the split side-bands, strong cooling at blue-detuning
and simultaneous heating and cooling resonances. 
Although other proposals also permit strong cooling rates,
 the multiple sidebands provide an exceptionally broad region of strong cooling,
 offering considerable robustness to experimental errors in the driving power, detunings and 
even the phase (a variation of $\phi$ of order 30\% will not appreciably
perturb the strong cooling).

\section{APPENDIX}
\subsection{Minimum phonon number}

From quantum perturbation theory we can show that $R_{n\to m}$, the rate of transition from state n to n+1 is:

\begin{equation}
R_{n \to n+1}=(n+1)\frac{\epsilon^2 \kappa_A}{\omega_M} \left( S_1(\omega_M)+S_2(\omega_M) \right)
\label{rateup}
\end{equation}

Similarly:

\begin{equation}
R_{n \to n-1}=n \ \frac{\epsilon^2 \kappa_A}{\omega_M} \left( S_1(-\omega_M)+S_2(-\omega_M) \right)
\label{ratedown}
\end{equation}

For $n>>1$, this gives the cooling rate of (Eq.\ref{gamma}).  However, with the
 exact expressions we can also show the equilibrium mean phonon number to be:

\begin{equation}
\bar{n}_{min}=\frac{S_1(-\omega_M)+S_2(-\omega_M)}{S_1(\omega_M)+S_2(\omega_M)-S_1(-\omega_M)-S_2(-\omega_M)}
\label{nmin}
\end{equation}

\begin{figure}[htb] 
\begin{center}
\includegraphics[height=5in]{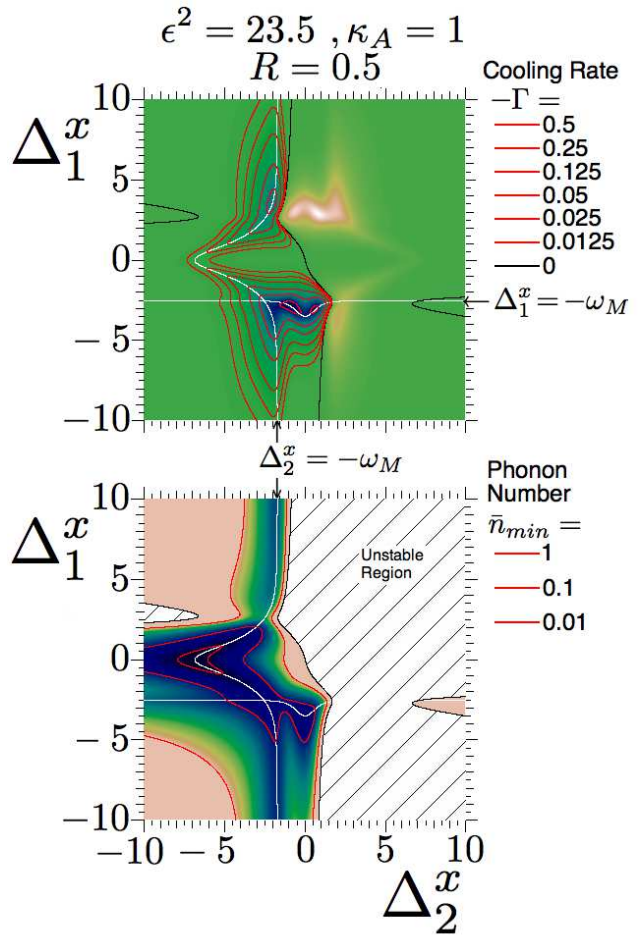} 
\end{center}
\caption{Colour online) maps of cooling rate and minimum phonon number for parameters $R=0.5$ equivalent to
Fig.\ref{Fig1}.  The cooling rate map shows information very
 similar to the shifted curves in Fig.\ref{Fig1}.
The white lines indicate the locus of the single field resonances $r1\pm$(where $\Delta_{1\pm}^x=\omega_M$
and $r2\pm$, where ($\Delta_{2\pm}^x=\omega_M$). Even at the Double Resonance (where the two white lines intersect),
there is still a contribution from the partly resolved third sideband of $r1+$, giving a very broad strong-cooling region
(dark blue). The corresponding broad strong-heating region is also seen for positive $\Delta_1^x$.
{\em NB:} 
The axis correspond to corrected detuning scaled to $A=\kappa$, $\Delta_{1,2}^x$, not the experimental $\delta_{1,2}$
of Fig.\ref{Fig1}. $ R=0.5$ maximum cooling is $\Gamma=-0.59$.}
\label{Fig4} 
\end{figure}

\begin{figure}[htb] 
\begin{center}
 \includegraphics[height=5in]{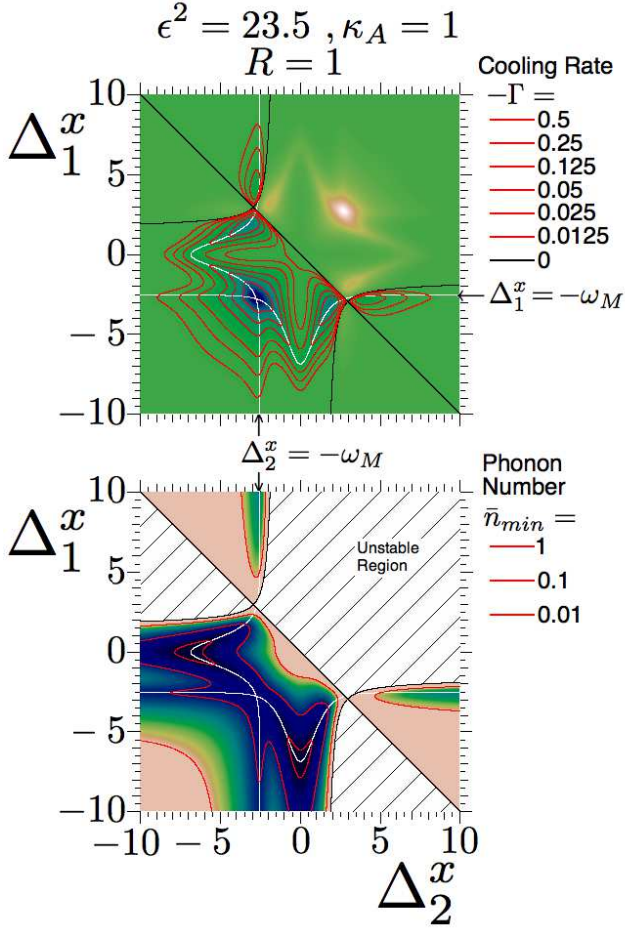} 
\end{center}
\caption{As for Fig.\ref{Fig4} but for $R=1$. The fields are equivalent, thus the 
cooling/heating maps have a high degree of reflection symmetry.
 The white lines indicate single field resonances. double resonance occurs where the two white lines intersect.
In this case, the splitting between $r1\pm$ is larger so the double-resonance involves only
$r1-$ and $r2-$.
The $ R=1.0$ max cooling is $\Gamma=-0.95$, $\Delta_{1}^x=\Delta_{2}^x=-2.72$.  }
\label{Fig5} 
\end{figure}

The parameters for strongest cooling do not necessarily yield the minimum phonon numbers,
 because the rate depends upon the difference between the optical heating
 and cooling, while the phonon number depends upon the ratio between the two.
In Figs.\ref{Fig4} and \ref{Fig5} we present colour maps comparing the cooling
and minimum phonon numbers for both $R=0.5$ and $R=1$ respectively.
Fig.\ref{Fig4} corresponds to the same parameters as Fig.\ref{Fig1}. It shows the
broad strong cooling region corresponding to the three distinct cooling resonances
$r2-, r1\pm$ which are only partially resolved. The result is a strong cooling
region of about 1MHz width, providing the advantage of a strong
cooling regime $\Gamma \sim \kappa$ insensitive to experimental detunings.
For Fig.\ref{Fig5} in contrast, the map has a 
high degree of symmetry, since  the role of the modes is interchangeable; the larger $R$
means the splitting is larger, so $r1\pm$ are fully resolved at the double resonance: thus
only $r2-,r1-$ contribute to the maximal cooling region. Nevertheless, this is the point
where the strongest cooling ($\Gamma= 0.95 \kappa_A$) is obtained.

The figures show that the largest cooling rates are found in the symmetric double resonance regime
(the double resonance for $R=1$),
 the phonon occupancy is slightly lower in the usually analysed \cite{Zoller} single-field regime 
where one strong field, at near zero detuning,
 traps while the other field cools.  Nevertheless, the mean equilibrium occupancy is very small
in both cases (less than a tenth of a phonon). The above analysis of minimum phonon occupancy
has not considered other sources of heating (photon scattering, background gas collisions).
The effect of these other heating rates has been analysed in \cite{Zoller} where ground
state cooling is shown to be achievable for optimal cooling rates ($\Gamma \sim \kappa$).
Thus the 1-2 order of magnitude increase in cooling in the double-resonance region means
that this is the most favourable regime.

 \subsection{r2 and single field cooling}

The single-field cooling rates are  obtained from
Eq.\ref{gamma} by taking $x_0 \simeq 0$,  $\Delta_1^x=0$ and $\Delta_2^x=-\omega_M$.

Hence,
\begin{equation}
\Gamma/2=\frac{\epsilon^2 \kappa_A}{2\omega_M}|\alpha_2|^2
\left[\frac{1}{[\omega_M-\Delta_2^x]^2 + \kappa_A^2}-\frac{1}{[\omega_M+\Delta_2^x]^2 + \kappa_A^2}\right]
\label{Chang}
\end{equation}

We can obtain the precise form given previously \cite{Zoller} 
 if we replace $|\alpha_2|^2$ using Eq.\ref{aeq}:
$\frac{|\alpha_2|^2}{|\alpha_1|^2} \simeq 2x_0 \simeq \frac{\kappa_A^2}{(\Delta_2+A/2)^2 + \kappa_A^2}$
and revert to unscaled units.

The above, in principle, requires a full numerical solution of  the equations:

\begin{eqnarray}
\alpha_{1} &=&  \left[\kappa_A - i\Delta_{1}^x \right]^{-1} \ ; \ 
\alpha_{2}  =  R\left[\kappa_A - i\Delta_{2}^x \right]^{-1} \nonumber\\
  \tan 2x_0 &=& {|\alpha_2|^2}/{|\alpha_1|^2} ,
\label{aeq}
\end{eqnarray}

to find equilibrium positions $x_0$ and fields. However we can obtain a good
approximation in closed form using experimental parameters (driving power, detunings).
Hence,
\begin{eqnarray}
\omega_M^2=  \frac{2\epsilon^2}{\kappa_A^2}
\end{eqnarray}
For resonant cooling $S_2(-\omega_M)\gg S_2(+\omega_M)$ and cooling/heating by field 1 is
negligible. So, 
\begin{equation}
-\Gamma\approx \frac{\epsilon^2|\alpha_2|^2}{\omega_M\kappa_A} 
\end{equation}
and $|\alpha_2|^2 = \frac{R^2}{\omega_M^2+\kappa_A^2}$, hence the Single Resonance
(SR) cooling rate becomes:

\begin{equation}
-\Gamma_{SR}\approx \frac{R^2\epsilon\kappa_A^2}{\sqrt{2}(2\epsilon^2+\kappa_A^4)}
\label{SR}
\end{equation}
{\em NB: this is a scaled cooling rate thus given in units of $A$}.
This gives a maximum cooling rate if $\epsilon=\kappa_A^2/\sqrt{2}$ where
 \begin{equation}
-\Gamma_{SR}\approx \frac{R^2}{4}
\label{SRR}
\end{equation}

It is worth noting that this maximum is independent of $\kappa_A$: it depends only on $R$. (Within the underlying assumption that $\omega_M>>\kappa_A$)
Even for $R=1/2$ , the maximum cooling is $\Gamma_{SR}/2\sim 1/32$ far from optimal
 $\Gamma_{opt}/2 \sim 1$.

As the driving is increased, if $2\epsilon^2 \gg \kappa_A^4$, 
 \begin{equation}
-\Gamma_{SR}\sim \frac{R^2 \kappa_A^2}{2\sqrt{2} \epsilon} \propto 1/\epsilon
\end{equation}
Thus the single resonance cooling rate falls off quite rapidly as the driving
amplitude is increased: the optimal cooling cannot be attained 
by increasing the driving amplitude.

\subsection{$r2-$ and $r1-$ overlap: double-resonance cooling}

As illustrated in Fig.\ref{Fig1}, the resonances $r2-$ and $r1-$ never
actually coincide: they undergo something reminiscent of an
avoided crossing (recall that Fig.\ref{Fig1} is a map of the classical
cooling, not of underlying quantum eigenvalues. Nonetheless,
the similarity between the classical linearisation and the effective Hamiltonian
for the quantum fluctuations has a very similar structure.

The double-field cooling rates for the region of closest 
approach are estimated from 
Eq.\ref{gamma} by assuming that the two resonances actually cross,
in other words both fields are, to a good approximation, simultaneously
resonant.

The first task is to identify the $\epsilon$-dependent 
pair of detunings for which:
\begin{equation}
-\omega_M(\Delta_1,\Delta_2) \approx \Delta^x_1 \approx \Delta^x_2
\end{equation}

One can search numerically for detunings which give near-simultaneous resonances.
However, for moderate $\sin 2x_0 \simeq \tan 2x_0 \simeq 2x_0$, a closed form can be obtained.
Provided $R$ is small the double resonance still falls in the small angle regime  (including our Fig.\ref{Fig1}).
In this case, From Eq.\ref{aeq}:
\begin{equation}
 2x_0 \simeq R^2 \frac{(\Delta_1^x)^2+\kappa_A^2}{(\Delta_2^x)^2 + \kappa_A^2}= R^2
\end{equation}
so if $\Delta_1^x=\Delta_2^x=-\omega_M$ we obtain $x_0 \simeq R^2/2$.

The mechanical frequency:
\begin{eqnarray}
\omega_M^2= 2\epsilon^2|\alpha_1|^2 = \frac{2\epsilon^2}{(\Delta_1^x)^2+\kappa_A^2}
=\frac{2\epsilon^2}{\omega_M^2+\kappa_A^2}
\label{omR2}
\end{eqnarray}
this means:
\begin{eqnarray}
\omega_M^2=  \frac{-\kappa_A^2}{2} + \frac{1}{2} \sqrt{ \kappa_A^4+ 8\epsilon^2}
\label{omm}
\end{eqnarray}


In this case, there are  contributions to cooling from both field 1 and field 2. Adding
them both,

\begin{equation}
-\Gamma_{DR}\approx \frac{\epsilon^2 (R^2+R^4)}{\kappa_A \omega_M (\omega_M^2+\kappa_A^2)}
\label{DR}
\end{equation}
where we will substitute Eq.\ref{omm} to evaluate the frequency $\omega_M$.
As ever, $\omega_M \gg \kappa_A$:
\begin{equation}
-\Gamma_{DR}\approx \frac{R^2\epsilon^{1/2}}{2^{3/4}\kappa_A}
\label{DR1}
\end{equation}

Eqs.\ref{DR} and \ref{DR1} should be contrasted with the behaviour of the singly resonant
cooling Eq.\ref{SRRE}. In the double-resonant case, the cooling increases, without limit,
as a function of $\epsilon$. For $R=1/2$ and $\epsilon^2\simeq 100$ (ie 8mW power),
$\Gamma \simeq 0.4 A$.
In Fig.\ref{Fig3} the approximate expressions Eqs.\ref{SRRE},\ref{DR},\ref{DR1} are shown
to give quite good agreement with cooling rates obtained from numerical
solution of the equations of motion, without linearisation.

The very large damping rates provided by the double-resonant regime have the
added advantage of relative insensitivity to the initial preparation. Although the analysis assumes small oscillations about equilibrium $x_0$, a resonator prepared at the antinode of the trapping field ($x=0$) is very  rapidly
pulled towards equilibrium and oscillates about $x=x_0$. Consequently there is no need to
provide any initial displacement. In addition, the multiple resonance
region shown in Fig.\ref{Fig1} is surprisingly robust to errors in the relative phase
between the two modes. In Fig.\ref{Fig4} we see that a phase error of order 20\%
makes relatively little difference (and may even enhance the cooling). 

\subsection{Symmetric double-resonance cooling}

The best regime for strong cooling is one in which the fields are equally strong
 and where both are resonantly, red detuned. In this region:
\begin{equation}
\omega_M = \sqrt{\frac{-\kappa_A^2}{2}+ \sqrt{\frac{\kappa_A^4}{4}+2\sqrt{2}\epsilon^2}}
\end{equation}
and:
\begin{equation}
\Gamma_{opt} = \frac{\epsilon^2\kappa_A}{\omega_M}\frac{1}{\kappa_A^2+\omega_M^2}
\left(\frac{1}{\kappa_A^2}-\frac{1}{\kappa_A^2+4\omega_M^2}\right)\\
\end{equation}

In the limit $\omega_M>>\kappa_A$:
\begin{equation}
\Gamma_{opt} =\frac{2^{-9/8}\sqrt{\epsilon}}{\omega_M}
\end{equation}

This is not simply a special case of Eq.\ref{DR1} because we have moved away from small $x_0$ into the regime $x_0=\pi/8$.  However the difference between Eq.\ref{DR1} and the above is less than a factor of 2.

\begin{figure}[htb] 
\begin{center}
\includegraphics[height=2.5in]{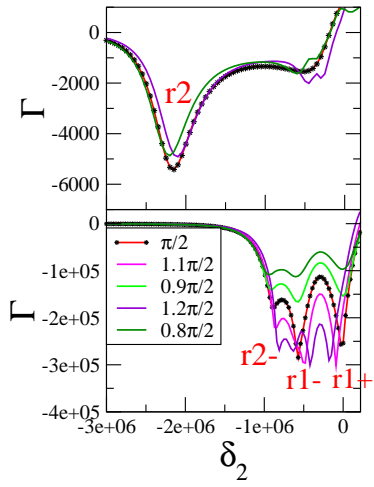} 
\end{center}
\caption{Shows that the strong cooling region where 
the split sidebands $r2-,r1-,r1+$ overlap is reasonably
insensitive to the phase difference between the two modes.
A variation of about 30\% about the usual dephasing of $\pi/2$
is tolerable (and can even give enhanced cooling).}
\label{Fig6} 
\end{figure}


\end{document}